\documentstyle[prd,aps,eqsecnum,floats,epsf]{revtex}
\draft
\begin{document}

%\preprint{MAD-NT-97-XX\\
%\hfill  DOE/ER/40561-3XX-INT97-XX-XXX}
\twocolumn[\hsize\textwidth\columnwidth\hsize\csname
@twocolumnfalse\endcsname
\title{
\hbox to\hsize{\large Submitted to Phys.~Rev.~D\hfil MAD-NT-97-09,  
DOE/ER/40561-347-INT97-00-186  }
\vskip2cm
Exact Solutions for Matter-Enhanced Neutrino Oscillations}
\author{A.~B. Balantekin\thanks{Electronic address:
        {\tt baha@nucth.physics.wisc.edu}}}
\address{Department of Physics, University of Wisconsin\\
         Madison, Wisconsin 53706 USA\thanks{Permanent Address}\\
         and\\
         Institute for Nuclear Theory, University of Washington, Box
         351550\\
         Seattle, WA 98195-1550 USA\\
         and\\
         Department of Astronomy, University of Washington, 
          Box 351580\\
       Seattle WA 98195-1580 USA}
\date{November 26, 1997}
\maketitle 

\begin{abstract}

The analogy between supersymmetric quantum mechanics and
matter-enhanced neutrino oscillations is exploited to obtain exact
solutions for a class of electron  density profiles. This
integrability condition is analogous to the shape-invariance in
supersymmetric quantum mechanics. This method seems to be the most
direct way to obtain the exact survival probabilities for a number of
density profiles of interest, such as linear and exponential density
profiles.  The resulting neutrino amplitudes can also be utilized as
comparison amplitudes for the uniform semiclassical treatment of
neutrino propagation in arbitrary electron density profiles.     

\end{abstract}

\pacs{14.60.Pq, 96.60.Jw, 26.65.+t}
\vskip2.2pc]

%%%%%%%%%%%%%%%%%%%%%%%%%%%%%%%%%%%%%%%%%%%%%%%%%%%%%%%%%%%%%%%%%%%%%%%%%%%%
%%%%%%%%%%%%%%%%%%%%%%%%%%%%%%%%%%%%%%%%%%%%%%%%%%%%%%%%%%%%%%%%%%%%%%%%%%%%

%\newpage

\narrowtext

\section{Introduction}

Matter enhanced oscillations of neutrinos via the
Mikheyev, Smirnov, Wolfenstein (MSW) mechanism \cite{msw} is of
considerable current interest. MSW mechanism may account for the
deficit of solar neutrinos \cite{sun}. Matter-enhanced neutrino
oscillations may also play an important role in neutrino propagation
through the core-collapse supernovae \cite{SN}. 

The equations of motion for the neutrinos in the MSW problem can be
solved by direct numerical integration, which must be repeated many
times when a broad range of mixing parameters are considered. This
often is not very convenient; consequently various approximations are
widely used. Exact or approximate analytic results allow a greater
understanding of the effects of parameter changes.   Exact solutions
of the MSW equations for a number of density profiles are available in
the literature. Landau-Zener method \cite{lz} is applicable to the
level-crossing phenomenon inherent in matter-enhanced neutrino
oscillations \cite{lzmsw,wick1}. The standard Landau-Zener method is
exact a linear density profile. Exact solutions for an exponential
density profile \cite{petcov1,toshev,wick2} or for a profile in the
form $\tanh x$ \cite{notz} are also given in the literature. 
 
The analogy between supersymmetric quantum mechanics and
matter-enhanced neutrino oscillations was pointed out in
Ref. \cite{baha1}. Indeed this analogy was used to obtain a
semiclassical approximate expression for the hopping probability
\cite{baha1,baha2}. The aim of the current paper is to exploit this
analogy further to obtain exact solutions for a class of electron
density profiles. The integrability condition we investigate is
analogous to the integrability condition called shape-invariance
\cite{genden} in supersymmetric quantum mechanics. 

for the purpose of establishing notation the MSW effect is outlined
and salient formulas are given in Section II. In Section III the
consequences of the  shape-invariance of electron density profile are
discussed, an exact expression for the electron neutrino amplitude for
such density profiles is derived and its asymptotic limits are
calculated. Several examples are given in Section IV and existing
results in the literature are shown to follow from the formalism
developed in the  previous section. In Section V, the shape-invariance
condition is shown to be a suitable first step for obtaining
approximate expressions for the neutrino survival probability. Section
VI includes a brief discussion of the results. 

\section{Outline of the MSW Effect}

The evolution of flavor eigenstates in matter is governed by the
equation 
\begin{equation}
i\hbar \frac{\partial}{\partial x} \left[\begin{array}{cc} \Psi_e(x)
\\ \\ \Psi_{\mu}(x) \end{array}\right] = \left[\begin{array}{cc}
\varphi(x) & \sqrt{\Lambda} \\ \\ \sqrt{\Lambda} & -\varphi(x)
\end{array}\right]
\left[\begin{array}{cc} \Psi_e(x) \\ \\ \Psi_{\mu}(x)
  \end{array}\right]\,, 
\label{1}
\end{equation}
where we defined 
\begin{equation}
  \label{2}
  \varphi(x) = \frac{1}{4 E} \left( 2 \sqrt{2}\ G_F N_e(x) E -  \delta
m^2 \cos{2\theta_v} \right)
\end{equation}
and
\begin{equation}
  \label{3}
  \sqrt{\Lambda} = \frac{\delta m^2}{4 E}\sin{2\theta_v}.
\end{equation}
In these equations   $\delta m^2 \equiv m_2^2 - m_1^2$ is the vacuum
mass-squared splitting, $\theta_v$ is the vacuum mixing angle,  $G_F$
is the Fermi  constant and $N_e(x)$ is the number density of electrons
in the medium.

By making a change of basis 
\begin{equation}
  \label{4}
  \left[\begin{array}{cc} \Psi_1(x) \\ \\ \Psi_2(x) \end{array}\right]
= \left[\begin{array}{cc} \cos{\theta(x)} & -\sin{\theta(x)} \\ \\
\sin{\theta(x)} & \cos{\theta(x)}
\end{array}\right]
\left[\begin{array}{cc} \Psi_e(x) \\ \\ \Psi_{\mu}(x)
\end{array}\right]\,,
\end{equation}
the flavor-basis Hamiltonian of Eq.~(\ref{1}) can be instantaneously
diagonalized. The matter mixing angle is defined via 
\begin{equation}
\sin{2\theta(x)} = \frac{\sqrt{\Lambda}}{\sqrt{\Lambda + \varphi^2(x)}}
\label{5}
\end{equation}
and
\begin{equation}
\cos{2\theta(x)} =  \frac{-\varphi(x)}{\sqrt{\Lambda +
\varphi^2(x)}}\,.
\label{6}
\end{equation}
In the adiabatic basis the evolution equation takes the form 
\begin{eqnarray}
i&\hbar& \frac{\partial}{\partial x} \left[\begin{array}{cc} \Psi_1(x)
\\ \\ \Psi_2(x) \end{array}\right] \nonumber \\ &=& 
\left[\begin{array}{cc}
-\sqrt{\Lambda + \varphi^2(x)} & -i \hbar \theta'(x) \\ \\ i \hbar
\theta'(x) & \sqrt{\Lambda + \varphi^2(x)}
\end{array}\right]
\left[\begin{array}{cc} \Psi_1(x) \\ \\ \Psi_2(x)
  \end{array}\right]\,, 
\label{7}
\end{eqnarray}
where prime denotes derivative with respect to $x$.  Since the
$2\times2$ ``Hamiltonian'' in Eq. (\ref{7}) is an element of the
$SU(2)$ algebra, the resulting time-evolution operator is an element
of the $SU(2)$ group. Hence it can be written in the form
\begin{equation}
  \label{7aa}
 U =  \left[\begin{array}{cc} \Psi_1(x)&  - \Psi_2^*(x) \\ \\
\Psi_2(x) & \Psi_1^*(x)
\end{array}\right] \,,
\end{equation}
where $\Psi_1(x)$ and $\Psi_2(x)$ are solutions of Eq. (\ref{7}) with
the initial conditions  $\Psi_1(x_i)=1$ and $\Psi_2(x_i)=0$. 

To calculate the electron neutrino survival probability  Eq. (\ref{1})
needs to be solved with the initial conditions $\Psi_e=1$ and
$\Psi_{\mu}=0$. Using Eq. (\ref{7aa})  the general solution satisfying
these initial conditions can be written as
\begin{eqnarray}
  \label{7a}
  \Psi_e (x) &=& \cos{\theta(x)} [ \cos{\theta_i} \Psi_1(x) -
\sin{\theta_i} \Psi_2^* (x)] \nonumber \\ &+& 
\sin{\theta(x)} [ \cos{\theta_i}
\Psi_2(x) +  \sin{\theta_i} \Psi_1^* (x)],
\end{eqnarray}
where  $\theta_i$ is the initial matter angle. Once the neutrinos
leave the dense matter (e.g. the Sun), the solutions of Eq. (\ref{7})
are particularly simple. Inserting these into Eq. (\ref{7a}) we obtain
the electron neutrino amplitude at a distance $L$ from the solar
surface to be 
\begin{eqnarray}
  \label{7b}
   &\Psi_e& (L) = \cos{\theta_v} [ \cos{\theta_i} \Psi_{1,(S)} -
\sin{\theta_i} \Psi_{2,(S)}^* ] \exp{\left(i\frac{\delta
m^2}{4E}L\right)} \nonumber \\ &+& \sin{\theta_v} [ \cos{\theta_i}
\Psi_{2,(S)} - \sin{\theta_i} \Psi_{1,(S)}^* ]
\exp{\left(-i\frac{\delta m^2}{4E}L\right)},
\end{eqnarray}
where $\Psi_{1,(S)}$ and $\Psi_{2,(S)}$ are the values of  $\Psi_1(x)$
and $\Psi_2(x)$ on the solar surface.  The electron neutrino survival
probability averaged over the detector position, $L$, is then given by 
\begin{eqnarray}
  \label{7c}
  &P&(\nu_e \rightarrow \nu_e) = \langle |\Psi_e (L)|^2
  \rangle_L\nonumber \\ &=&
\frac{1}{2} + \frac{1}{2} \cos{2\theta_v} \cos{2\theta_i}  \left( 1
- 2 |\Psi_{2,(S)}|^2 \right) \nonumber \\ &-& \frac{1}{2}
\cos{2\theta_v} \sin{2\theta_i} \left( \Psi_{1,(S)} \Psi_{2,(S)} +
\Psi_{1,(S)}^* \Psi_{2,(S)}^* \right)\,.
\end{eqnarray}
If the initial density is rather large, then $\cos{2\theta_i} \sim -1$
and  $\sin{2\theta_i} \sim 0$ and the last term in Eq. (\ref{7c}) is
very  small.  Different neutrinos arriving the detector carry
different phases if they are produced over an extended source. Even if
the initial matter density is not very large, averaging over the
source position makes the last term very small as these phases average
to zero. The completely averaged result for the electron neutrino
survival probability is then given by 
\begin{equation}
  \label{7d}
   P(\nu_e \rightarrow \nu_e) = \frac{1}{2} + \frac{1}{2}
\cos{2\theta_v} \langle \cos{2\theta_i} \rangle_{\rm source}  \left( 1
- 2 P_{\rm hop} \right), 
\end{equation}
where the hopping probability is 
\begin{equation}
  \label{7e}
  P_{\rm hop} = |\Psi_{2,(S)}|^2, 
\end{equation}
obtained by solving Eq. (\ref{7}) with the initial conditions
$\Psi_1(x_i)=1$ and $\Psi_2(x_i)=0$. 

\section{Shape-Invariance of the Electron Density Profile}

The coupled first-order equations for the flavor-basis wave functions
can be decoupled to yield a second order equation for only the
electron neutrino propagation 
\begin{equation}
-\hbar^2\frac{\partial^2 \Psi_e(x)}{\partial x^2} - \left[\Lambda +
\varphi^2(x) + i\hbar\varphi'(x)\right]\Psi_e(x) = 0.
\label{8}
\end{equation}
Introducing the operators
\begin{eqnarray}
  \label{9}
  {\hat A}_- &=& i\hbar \frac{\partial}{\partial x} - \varphi(x) \,,
\nonumber \\ {\hat A}_+ &=& i\hbar \frac{\partial}{\partial x} +
\varphi(x) \,,
\end{eqnarray}
Eq. (\ref{1}) takes the form 
\begin{eqnarray}
  \label{10}
  {\hat A}_- \Psi_e(x) &=& \sqrt{\Lambda} \Psi_{\mu}(x)\,, \nonumber
\\ {\hat A}_+ \Psi_{\mu}(x) &=& \sqrt{\Lambda} \Psi_e(x)\,,
\end{eqnarray}
and Eq. (\ref{8}) can be compactly expressed as
\begin{equation}
   {\hat A}_+ {\hat A}_- \Psi_e(x) = \Lambda \Psi_e(x).
  \label{11}
\end{equation}
In these equations we take $\varphi(x), {\hat A}_+$, and ${\hat A}_-$
a function of a set of parameters $a$ that specify space-independent
properties of the electron density. 

The analogy between Eq. (\ref{8}) and supersymmetric quantum mechanics
was pointed out some time ago \cite{baha1}. In a parallel development,
in the context of supersymmetric quantum mechanics 
it was shown that a subset of the potentials for which the
Schr\"odinger equations are exactly solvable share an integrability
condition called shape invariance \cite{genden}. In this paper we
consider a generalization of the shape-invariance condition to
electron densities. We call a given electron density to be
shape-invariant if a change of parameters satisfies the condition 
\begin{eqnarray}
  \label{12}
  - \varphi^2(x;a_1) &+& i\hbar\varphi'(x;a_1)\nonumber \\ &=& 
  - \varphi^2(x;a_2) - i\hbar\varphi'(x;a_2) + R(a_1) \,,
\end{eqnarray}
where $a_2$ is a function of $a_1$, and the remainder $R(a_1)$ is
independent of $x$. As we illustrate in the forthcoming sections a
number of electron density profiles satisfy this condition. 
Here we take $a_1$ to be the original set of parameters specifying the
given electron density profile (hence real). The new set of parameters
$a_2$ are, in general, complex. The shape invariance condition of Eq. 
(\ref{12}) can be rewritten in terms of the operators defined in
Eq. (\ref{9}) 
\begin{equation}
  \label{13}
  {\hat A}_-(a_1) {\hat A}_+(a_1) = 
  {\hat A}_+(a_2) {\hat A}_-(a_2) + R(a_1). 
\end{equation}
We assume that replacing $a_1$
with $a_2$ in a given operator can be achieved with a similarity
transformation: 
\begin{equation}
  \label{14}
  \hat{T}(a_1) {\cal O}(a_1) \hat{T}^{-1}(a_1) = {\cal O}(a_2).
\end{equation}
If the
parameters $a_1$ and $a_2$ are related by a translation,  
\begin{equation}
  \label{15} 
  a_2 = a_1 + \eta, 
\end{equation}
then the the operator $\hat{T}(a_1)$ of
Eq. (\ref{14}) is simply given by
\begin{equation}
  \label{16}
  \hat{T}(a_1) = \exp \left( \eta \frac{\partial}{\partial a_1}
  \right).
\end{equation}
One should emphasize that the discussion in this article is valid for
any relationship between $a_1$ and $a_2$ and is not limited to that
given in Eq. (\ref{15}). In the rest of this paper $\hbar$ is set to
one for typographical convenience, except when we discuss the
semiclassical limit. 

\subsection{Exact Solutions}

To solve the MSW equations we first introduce the operators
\cite{baha3} 
\begin{eqnarray}
  \label{17}
  \hat{B}_+ &=& \hat{A}_+(a_1) \hat{T} (a_1) \nonumber \\
 \hat{B}_-  &=& \hat{T}^{-1}(a_1) \hat{A}_- (a_1) \,. 
\end{eqnarray}
Using Eq. (\ref{13}) one can easily prove the commutation relation: 
\begin{equation}
  \label{18}
  [ \hat{B}_- , \hat{B}_+ ] = R(a_0) ,
\end{equation}
where we used the identity 
\begin{equation}
  \label{19}
  R(a_n) = \hat{T}(a_1) R(a_{n-1})\hat{T}^{-1}(a_1),    
\end{equation}
valid for any $n$, to define $a_0$. 
One can also prove by induction two additional commutation relations 
\begin{equation}
  \label{20}
   [\hat{B}_+ \hat{B}_- , \hat{B}_+^n ] = (R(a_1)+R(a_2)+ \cdot \cdot 
+ R(a_n))  \hat{B}_+^n , 
\end{equation}
and 
\begin{equation}
  \label{21}
  [\hat{B}_+ \hat{B}_- , \hat{B}_-^{-n} ] = (R(a_1)+R(a_2)+ \cdot \cdot 
+ R(a_n))  \hat{B}_-^{-n} \,.  
\end{equation}

Using the operators introduced in Eq. (\ref{17}), Eq. (\ref{11}) can
be rewritten as 
\begin{equation}
  \label{22}
  {\hat B}_+ {\hat B}_- \Psi_e(x) = \Lambda \Psi_e(x).
\end{equation}
Eqs. (\ref{20}) and (\ref{21}) imply that ${\hat B}_+$ and ${\hat B}_-$
can be used as ladder operators to solve Eq. (\ref{22}). To this  end 
we introduce $\Psi_-^{(0)}$ as the solution of the equation 
\begin{equation}
  \label{23}
   {\hat A}_-(a_1) \Psi_-^{(0)} = 0 = {\hat B}_- \Psi_-^{(0)} ,
\end{equation}
which implies 
\begin{equation}
  \label{24}
  \Psi_-^{(0)}  \sim \exp{\left( - i \int \varphi(x;a_1) dx
    \right)}. 
\end{equation}
If the function 
\begin{equation}
  \label{25}
  f(n) = \sum_{k=1}^{n} R(a_k) 
\end{equation}
can be analytically continued so that the condition
\begin{equation}
  \label{26}
  f(\mu) = \Lambda 
\end{equation}
is satisfied for a particular (in general, complex) value of $\mu$,
then Eq. (\ref{20}) implies that one solution of Eq. (\ref{22}) is 
${\hat B}_+^{\mu} \Psi_-^{(0)}$. Similarly if $\Psi_+^{(0)}$ satisfies
the equation
\begin{equation}
  \label{27}
  {\hat B}_+ \Psi_+^{(0)} = 0, 
\end{equation}
which implies that
\begin{equation}
  \label{28}
  {\hat T}(a_1) \Psi_+^{(0)}  \sim \exp{\left( + i \int \varphi(x;a_1) dx
    \right)} ,
\end{equation}
or
\begin{equation}
  \label{29}
   \Psi_+^{(0)}  \sim \exp{\left( + i \int \varphi(x;a_0) dx \right)}\,,
\end{equation}
then Eq. (\ref{21}) implies that a second solution of Eq. (\ref{22}) 
is given by
${\hat B}_-^{-\mu-1} \Psi_+^{(0)}$. Hence for shape invariant electron
densities the exact electron neutrino amplitude can be written as
\begin{eqnarray}
  \label{30}
  \Psi_e (x) &=& \beta {\hat B}_+^{\mu} \exp{\left( -  i \int
      \varphi(x;a_1) dx \right)} \nonumber \\ &+& 
\gamma {\hat B}_-^{-\mu-1} 
      \exp{\left( + i \int \varphi(x;a_0) dx \right)}, 
\end{eqnarray}
where $\beta$ and $\gamma$ are constants to be determined by the
initial conditions. 

\subsection{Asymptotic Results}

To evaluate Eq. (\ref{30}) asymptotically we first note that using Eq.
(\ref{17}) the two terms in  Eq. (\ref{30}) can be cast in the forms
\begin{eqnarray}
  \label{31}
  &{\hat B}_+^n& \Psi_-^{(0)} = \nonumber \\  
&{\hat A}_+(a_1)& {\hat A}_+(a_2) \cdot
\cdot {\hat A}_+(a_n) \exp{\left( - i \int \varphi(x;a_{n+1}) dx
\right)}, 
\end{eqnarray}
and
\begin{eqnarray}
  \label{32}
  &{\hat B}_-^{-n}& \Psi_+^{(0)} = \nonumber \\ &{\hat A}_-^{-1}(a_1)&
 {\hat
A}_-^{-1}(a_2) \cdot \cdot {\hat A}_-^{-1}(a_n) \exp{\left( + i \int
\varphi(x;a_n) dx \right)}.
\end{eqnarray}   
We consider two different limits: i) $\varphi(x)$ is constant (i.e.,
either the electron density is constant, or the neutrino is traveling
through the vacuum); and ii) $\varphi(x) \rightarrow \pm \infty$. In
the former limit the commutator 
\begin{equation}
  \label{33}
 [ \frac{\partial}{\partial x} , \varphi(x;a_n) ] = \varphi'(x;a_n)
\end{equation}
vanishes. In the latter case this commutator can be ignored as
\begin{eqnarray}
  \label{34}
  &\varphi(x;a_n)& \varphi(x;a_m) + i \varphi'(x;a_n)\nonumber \\
  &=&
\varphi(x;a_n) \varphi(x;a_m) \left( 1 + i
\frac{\varphi'(x;a_n)}{\varphi(x;a_n) \varphi(x;a_m)} \right)
\nonumber \\ &\rightarrow&  \varphi(x;a_n) \varphi(x;a_m) 
\end{eqnarray}
provided that $\varphi'(x;a_n)/\varphi(x;a_n)$ remains finite (which
is the case for all realistic density profiles). Hence it both limits
we can write Eqs. (\ref{31}) and (\ref{32}) as
\begin{eqnarray}
  \label{35}
   &{\hat B}_+^n& \Psi_-^{(0)} = (\varphi(a_{n+1}) + \varphi(a_1))
(\varphi(a_{n+1}) + \varphi(a_2)) \nonumber \\  &\cdots& \, \times
(\varphi(a_{n+1}) + \varphi(a_n))  \exp{\left( - i \int
\varphi(x;a_{n+1}) dx \right)}, 
\end{eqnarray}
and 
\begin{eqnarray}
  \label{36}
   &{\hat B}_-^{-n}& \Psi_+^{(0)} = (- \varphi(a_{n}) -
\varphi(a_1))^{-1} (- \varphi(a_{n}) - \varphi(a_2))^{-1} \nonumber \\
&\cdots&  \, \times (- \varphi(a_{n}) - \varphi(a_n))^{-1} \exp{\left(
+ i \int \varphi(x;a_n) dx \right)}.
\end{eqnarray}
In both of these equations the quantity $\varphi(a_{n})$ stands for
$\varphi(x;a_{n})$. 

If the density profile satisfies the condition 
\begin{equation}
  \label{37}
  \varphi(x;a_{n}) = \varphi(x;a_1) + (n-1) \zeta (x), 
\end{equation}
then these asymptotic equations can be cast in a form suitable for
analytic continuation :
\begin{equation}
  \label{38}
  {\hat B}_+^n \Psi_-^{(0)} = \zeta^n  \frac{\Gamma(2n + 2
\varphi(a_1)/\zeta)}{\Gamma(n + 2 \varphi(a_1)/\zeta)}  \exp{\left( -
i \int \varphi(x;a_{n+1}) dx \right)}, 
\end{equation}
and 
\begin{eqnarray}
  \label{39}
  {\hat B}_-^{-n} \Psi_+^{(0)} &=& (-\zeta)^n \frac{\Gamma(n - 1 + 2
\varphi(a_1)/\zeta)}{\Gamma(2n - 1 + 2 \varphi(a_1)/\zeta)} 
\nonumber \\ &\times&
\exp{\left( + i \int \varphi(x;a_n) dx \right)}.
\end{eqnarray} 
The condition in Eq. (\ref{37}) is satisfied for a number of density
profiles as illustrated in the next section. If this condition is not
satisfied, the analytic continuation may still be done, but will be
more complicated. Whenever the condition in Eq. (\ref{37}) is
satisfied, in the asymptotic limits the electron neutrino amplitude
can be written as 
\begin{eqnarray}
  \label{40}
   \Psi_e &=& \beta \zeta^{\mu} \frac{\Gamma(2\mu + 2
\varphi(a_1)/\zeta)}{\Gamma(\mu + 2 \varphi(a_1)/\zeta)}  \exp{\left(
- i \int \varphi(x; a_{\mu +1}) dx \right)} \nonumber \\ &+& \gamma
(-\zeta)^{-\mu-1} \frac{\Gamma(\mu + 2 \varphi(a_1)/\zeta)}{\Gamma(2
\mu +1  + 2 \varphi(a_1)/\zeta)} \nonumber \\ &\times&
 \exp{\left( + i \int
\varphi(x;a_{\mu+1}) dx \right)}.
\end{eqnarray}

Also using the identity
\begin{equation}
  \label{40d}
  \lim_{y \rightarrow \pm \infty} \frac{1}{y^{\mu}} \frac{\Gamma(2\mu
+ y)}{\Gamma(\mu + y)} = 1, 
\end{equation}
the asymptotic value of Eq. (\ref{40}) for $\varphi \rightarrow \pm
\infty$ is found to be
\begin{eqnarray}
  \label{40e}
   \Psi_e &=& \beta \left(2 \varphi(a_1) \right)^{\mu} \exp{\left( - i
\int \varphi(x;a_{\mu+1}) dx \right)} \nonumber \\ &+&
 \gamma \left(-2 \varphi(a_1)
\right)^{-\mu-1}  \exp{\left( + i \int \varphi(x;a_{\mu +1}) dx
\right)}\,.
\end{eqnarray}
 
The hopping probability can be easily obtained from these results as
we illustrate in the next section. 

\section{Examples}

\subsection{Linear density profile}

For the linear density profile
\begin{equation}
  \label{41}
  N_e(x) = N_0 - N_0' (x-x_R),
\end{equation}
where $N_0$ is the resonant density: 
\begin{equation}
  \label{42}
  2 \sqrt{2}\ G_F N_0 E =  \delta m^2 \cos{2\theta_v},
\end{equation}
we can write
\begin{equation}
  \label{43}
  \varphi(x) = - \frac{\delta m^2}{4E} \cos{2\theta_v}
\frac{N_0'}{N_0} (x-x_R). 
\end{equation}
Adopting the ansatz $\varphi(x;a_n) = a_n (x-x_R)$, one finds that
Eq.(\ref{12}) is satisfied with 
\begin{equation}
  \label{44}
  a_1 = - \frac{\delta m^2}{4E} \cos{2\theta_v} \frac{N_0'}{N_0} = a_2
= a_3 \cdots,
\end{equation}
and
\begin{equation}
  \label{45}
  R(a_n) = 2i a_1 = -i \frac{\delta m^2}{2E} \cos{2\theta_v}
\frac{N_0'}{N_0},
\end{equation}
for all $n$ (i.e., $\zeta=0$ for a linear density profile in Eq.
(\ref{37}). The function $f(n)$ of Eq. (\ref{25}) is then 
\begin{equation}
  \label{46}
  f(n) = -i \left(\frac{\delta m^2}{2E} \cos{2\theta_v} \right) n
\end{equation}
and hence the solution of Eq.(\ref{26}) is given by
\begin{equation}
  \label{47}
  \mu = \frac{i}{2} \Omega,
\end{equation}
where we defined
\begin{equation}
  \label{48}
   \Omega = \frac{\delta m^2}{4E}
\frac{\sin^2{2\theta_v}}{\cos{2\theta_v}} \frac{N_0}{N_0'}. 
\end{equation}

Linear density profile is special in the sense that both before and
after the MSW resonance its absolute value gets to be very
large. Hence Eq. (\ref{40e}) can be used to evaluate both asymptotic
limits. Initially we take $\varphi \rightarrow + \infty$ which gives 
\begin{equation}
  \label{49}
  \Psi_e (x)= \beta \left(2 \varphi(x) \right)^{i \Omega/2}
\exp{\left( - i \int_0^x \varphi(x_1) dx_1 \right)}. 
\end{equation}
Imposing the initial condition $\Psi(x_i)=1$ determines $\beta$ and
gives the final ($\varphi \rightarrow - \infty$) solution to be
\begin{equation}
  \label{50}
  \Psi_e (x)= \left( \frac{\varphi(x)}{\varphi(x_i)} \right)^{i
\Omega/ 2}  \exp{\left( - i \int_0^x \varphi(x_1) dx_1 \right)}. 
\end{equation}
Using Eqs. (\ref{5}) and (\ref{6}) one finds that the asymptotic
matter angles are
\begin{eqnarray}
  \label{50a}
  \varphi &\rightarrow& + \infty, \,\, \sin{\theta_i} =1,
\cos{\theta_i}=0, \nonumber \\ \varphi &\rightarrow& - \infty, \,\,
\sin{\theta_f} =0, \cos{\theta_f}=1,
\end{eqnarray}
Inserting Eq. (\ref{50a}) into Eq. (\ref{7a}) one finds that 
\begin{equation}
  \label{50b}
  \Psi_e (x_f) = - \Psi_2^* (x_f).
\end{equation}
Comparing Eqs. (\ref{50}) and (\ref{50b}) and using 
\begin{equation}
\lim_{x_i \rightarrow -\infty,x_f \rightarrow +\infty}
\frac{\varphi(x_f)}{\varphi(x_i)} = -1 = e^{i\pi} \nonumber
\end{equation} 
 we find that the hopping probability is given by
\begin{equation}
  \label{50c}
  P_{\rm hop} = |\Psi_2 (x_f)|^2 = \exp{(- \pi \Omega)} \,. 
\end{equation}

\subsection{Exponential Density Profile}

In this case we have
\begin{equation}
  \label{51}
  N_e(x) = N_0 e^{-\alpha(x-x_R)}, 
\end{equation}
where $N_0$ is the resonant density given in Eq. (\ref{42}),  we can
write
\begin{equation}
  \label{52}
  \varphi(x) =  \frac{\delta m^2}{4E} \cos{2\theta_v} \left(
e^{-\alpha(x-x_R)} - 1 \right). 
\end{equation} 
Adopting the ansatz
\begin{equation}
  \label{53}
   \varphi(x;a_n) =  \frac{\delta m^2}{4E} \cos{2\theta_v} \left(
e^{-\alpha(x-x_R)} - a_n \right)\,\,,a_1=1, 
\end{equation}
one can show that to satisfy Eq. (\ref{12}) we should require 
\begin{equation}
  \label{54}
  a_n = a_1 - \frac{i \alpha 4E}{\delta m^2 \cos{2\theta_v}} (n-1), 
\end{equation}
and
\begin{equation}
  \label{55}
  R(a_n) = - 2 i \alpha a_n \frac{\delta m^2}{4E} \cos{2\theta_v} -
\alpha^2 \,.
\end{equation}
Hence 
\begin{equation}
  \label{55a}
  \zeta = i \alpha 
\end{equation}
In this case there are two solutions of Eq. (\ref{26}) and the
appropriate solution giving the correct limiting behavior (that $\mu$
vanishes as $\Lambda$ does) is 
\begin{equation}
  \label{56}
  \mu = \frac{i}{\alpha} \left(\frac{\delta  m^2}{4E} \right) ( 1 -
\cos{2\theta_v}) . 
\end{equation}

Exponential density profile, like the linear one, can get to be very
large (and of course positive) before the MSW resonance point. Hence
initially the second term in Eq. (\ref{40e}) vanishes and the constant
$\beta$ can be fixed to be $ \beta = 2
(\varphi(x_i))^{-\mu}$. However, unlike the linear density, much after
the MSW resonance point the exponential density vanishes and one
should use the asymptotic form for the constant $\varphi$ derived in
the previous section. In this limit we have
\begin{eqnarray}
  \label{57}
  \varphi(x,a_1) &=& - \frac{\delta m^2}{4E} \cos{2\theta_v},
\nonumber\\  \varphi(x,a_{\mu+1}) &=& - \frac{\delta m^2}{4E}, 
\end{eqnarray} 
Inserting Eqs. (\ref{56}) and (\ref{57}) into Eq. (\ref{40}) we find
the final electron neutrino amplitude to be 
\begin{eqnarray}
  \label{58}
   \Psi_e (x_f) &=& \left( \frac{i\alpha}{2 \varphi(x_f)}\right)^{\mu}
\frac{\Gamma\left(\frac{i}{\alpha}\frac{\delta m^2}{2E}\right)}{\Gamma
\left( \frac{i}{\alpha}\frac{\delta m^2}{4E} (1+ \cos{2\theta_v})
\right)} \nonumber \\ &\times&
 \exp{\left( + i \frac{\delta m^2}{4E} x_f \right)} \nonumber
\\ &+& \gamma (-i\alpha)^{-\mu-1} \frac{\Gamma\left(
\frac{i}{\alpha}\frac{\delta m^2}{4E} (1+ \cos{2\theta_v})
\right)}{\Gamma\left(1+ \frac{i}{\alpha}\frac{\delta m^2}{2E} \right)}
\nonumber \\ &\times&
\exp{\left( - i  \frac{\delta m^2}{4E} x_f     \right)}.
\end{eqnarray}
The asymptotic matter angles are now given by
\begin{eqnarray}
  \label{59}
  \varphi &\rightarrow& + \infty, \,\, \sin{\theta_i} =1,
\cos{\theta_i}=0, \nonumber \\ \varphi &\rightarrow& 0, \,\,
\sin{\theta_f} = \sin{\theta_v},  \cos{\theta_f}= \cos{\theta_v}.
\end{eqnarray}
Inserting Eq. (\ref{59}) into Eq. (\ref{7b}) we get 
\begin{eqnarray}
  \label{60}
   \Psi_e (L) &=& - \cos{\theta_v} \Psi_{2,(s)}^*
\exp{\left(i\frac{\delta m^2}{4E}L\right)} \nonumber \\ &-& 
\sin{\theta_v}
\Psi_{1,(S)}^*   \exp{\left(-i\frac{\delta m^2}{4E}L\right)}\,.
\end{eqnarray}
Comparing Eqs. (\ref{58}) and (\ref{60}) one gets
\begin{eqnarray}
  \label{61}
  \Psi_{2,(s)}^* = &-& \frac{1}{\cos{\theta_v}} \left( \frac{\alpha}{2
\varphi(x_f)}\right)^{\mu}
\frac{\Gamma\left(\frac{i}{\alpha}\frac{\delta m^2}{2E}\right)}{\Gamma
\left( \frac{i}{\alpha}\frac{\delta m^2}{4E} (1+ \cos{2\theta_v})
\right)} \nonumber \\ &\times&  \exp{\left(- \frac{\pi}{\alpha}
\frac{\delta m^2}{8E} (1- \cos{2\theta_v})\right)} \,,
\end{eqnarray}
which, upon squaring, gives the hopping probability to be
\begin{equation}
  \label{62}
  P_{\rm hop} = \frac{e^{-\pi \delta (1 - \cos{2\theta_v})} - e^{-2
\pi \delta}}{1-e^{-2 \pi \delta}}, 
\end{equation}
where we defined
\begin{equation}
  \label{63}
  \delta = \frac{\delta m^2}{2E\alpha}. 
\end{equation}
Eq. (\ref{62}) was previously obtained by direct solution of the
differential equation \cite{wick2}. The method presented here is not
only more straightforward, but also explicitly demonstrates the role
of the boundary conditions. Indeed using the infinite Landau-Zener
method only provides the first term in the numerator of Eq. (\ref{61})
\cite{pizz}. To obtain the exact result, it is essential to take into
account the fact that the electron density goes to zero (not to
infinity) when the neutrinos move very far away from the MSW resonance
region as illustrated above. 

\section{Approximate Expressions}

The method presented here is useful not only to write down exact
expressions for the electron neutrino survival probabilities for
shape-invariant electron densities, but can also be used as a starting
point of several useful approximations. In particular, the
transformation in Eq. (\ref{14}) can be formally introduced without an
explicit reference to the parameters $a_n$. Suppose, for example, we
wish to satisfy Eq. (\ref{26}) with $n=1$ by introducing the function
$\xi(x)$ in Eq. (\ref{12})
\begin{equation}
  \label{64}
  \xi^2(x) - \varphi^2(x) + i \hbar \varphi'(x) + i \hbar \xi'(x) =
\Lambda. 
\end{equation}
In this section we write $\hbar$ explicitly in the equations to
emphasize the semiclassical nature of the approximations we
consider. By expanding the function $\xi(x)$ in powers of $\hbar$ 
\begin{equation}
  \label{65}
  \xi(x) = \xi_0(x) + \hbar \xi_1(x) + \hbar^2 \xi_2(x) + \cdots ,
\end{equation}
we find that Eq. (\ref{64}) can be solved by matching the powers of
$\hbar$: 
\begin{eqnarray}
  \label{66}
  \xi_0^2 &=& \varphi^2 + \Lambda, \nonumber \\ 2 \xi_0\xi_1 &=& -i
(\varphi' + \xi_0') \,.
\end{eqnarray}
Using Eqs. (\ref{30}), (\ref{31}), and (\ref{32}) the electron
neutrino amplitude in this approximation can be written as  
\begin{eqnarray}
  \label{67}
  &&\Psi_e(x) = \beta \left( i \hbar \frac{\partial}{\partial x} +
\varphi \right) \exp \left( - \frac{i}{\hbar} \int \xi(x) dx \right) 
 + \gamma
\nonumber \\ &\times&\left( i \hbar \frac{\partial}{\partial x} -
\varphi \right)^{-1} \left( i \hbar \frac{\partial}{\partial x} - \xi
\right)^{-1} \exp \left( + \frac{i}{\hbar} \int \xi(x) dx \right)\,. 
\end{eqnarray}
For a monotonically decreasing $\varphi$ we choose $\xi_0 = +
\sqrt{\varphi^2 + \Lambda}$. To evaluate Eq. (\ref{67}) in the lowest
order in $\hbar$, we note that in the exponentials one needs to
include $\xi_1$ (since $\hbar$'s cancel) whereas in the
pre-exponential factors it is sufficient to retain only
$\xi_0$. Inserting Eq. (\ref{66}) into Eq. (\ref{67}) and imposing the
initial conditions we obtain to lowest order in $\hbar$
\begin{eqnarray}
  \label{68}
  \Psi_e (x) &=& \frac{1}{2} T_-(0) T_-(x) \exp \left( +
\frac{i}{\hbar} \int \sqrt{\varphi^2(x) + \Lambda} dx \right)
\nonumber \\ &+& \frac{1}{2} T_+(0) T_+(x) \exp \left( -
\frac{i}{\hbar} \int \sqrt{\varphi^2(x) + \Lambda} dx \right)\,,
\end{eqnarray}
where we defined
\begin{equation}
  \label{69}
  T_{\pm}(x) = \left( 1 \pm \frac{\varphi(x)}{\sqrt{\varphi^2(x) +
\Lambda}} \right)^{1/2}. 
\end{equation}
Eq. (\ref{68}) is the electron neutrino amplitude in the standard
adiabatic approximation. The equivalence of the adiabatic and
primitive semiclassical approximations was previously proved in
Ref. \cite{baha1}. Here we demonstrated that it can be obtained by
solving a shape-invariance condition, valid for any electron density,
to lowest order in $\hbar$. 

Even though shape-invariant electron densities constitute a rather
restrictive class, one can use these exact solutions as comparison
solutions to obtain uniform semiclassical approximate solutions
\cite{mg,baha1,beac} for any given density that is not
shape-invariant. To achieve this goal, one starts with a
shape-invariant electron density profile for which the solution of the
``mapping'' equation 
\begin{equation}
  \label{70}
  \left( i \frac{\partial}{\partial S} + \xi(S) \right)  \left( i
\frac{\partial}{\partial S} - \xi(S) \right) \chi (S) = \Omega \chi(S)
\end{equation}
is known. In the next step, the variable $S$ in Eq. (\ref{70}) is
taken to a function of $x$ and  the solutions of Eq. (\ref{11}) are
written as
\begin{equation}
  \label{71}
  \Psi (x) = K(x) \chi (S(x))\,.
\end{equation}
By making the choice
\begin{equation}
  \label{72}
  K(x) = \frac{1}{\sqrt{S'(x)}}, 
\end{equation}
one can show that $\Psi$ given in Eq. (\ref{71}) satisfies
Eq. (\ref{11}) if the equality
\begin{eqnarray}
  \label{73}
  \hbar^2 \frac{K''}{K} &-& S'^2 \left( \Omega + \xi^2 + i \hbar
\frac{\partial\xi}{\partial S} \right) \nonumber \\ &+&\left( \Lambda
+ \varphi^2 + i \hbar \varphi' \right) = 0
\end{eqnarray}
is satisfied. Eq. (\ref{73}) is usually solved by expanding $S(x)$ in
powers of $\hbar$. Using a linear mapping density this mapping
technique was used in Refs. \cite{baha1} and \cite{baha2} to obtain an
approximate solution for the neutrino survival probability. It is
possible to generalize this result by using not the linear density,
but another shape-invariant mapping density, which most closely
resembles the actual electron density that the neutrinos travel
through. 

\section{Conclusions}

We showed that the analogy between supersymmetric quantum mechanics and
matter-enhanced neutrino oscillations can be used to obtain exact
expressions for the neutrino amplitudes traveling in 
a class of electron density profiles.  We explicitly worked out the
neutrino survival probability for two shape-invariant electron density
profiles that are most commonly used for studying matter-enhanced
neutrino oscillations, namely linear and exponential densities. In
addition to these two the $\tanh x$ and $1/(a+bx)$ density profiles
are also shape-invariant and the MSW equations for these density
profiles can be solved using the techniques described here. One should
emphasize that although the neutrino propagation in these density
profiles was studied before, the method described in this paper seems
to be more direct and to have pedagogical value as it explicitly 
illustrates the role of the boundary conditions. The resulting
neutrino amplitudes can also be utilized as comparison amplitudes for
the uniform semiclassical treatment of neutrino propagation in arbitrary
electron density profiles.

%%%%%%%%%%%%%%%%%%%%%%%%%%%%%%%%%%%%%%%%%%%%%%%%%%%%%%%%%%%%%%%%%%%%%%%%%%%%
%%%%%%%%%%%%%%%%%%%%%%%%%%%%%%%%%%%%%%%%%%%%%%%%%%%%%%%%%%%%%%%%%%%%%%%%%%%%

\section*{ACKNOWLEDGMENTS}

This work was supported in part by the U.S. National Science
Foundation Grant No.\ PHY-9605140 at the University of Wisconsin, and
in part by the University of Wisconsin Research Committee with funds
granted by the Wisconsin Alumni Research Foundation. I thank Institute
for Nuclear Theory and Department of Astronomy  at the University of
Washington for their hospitality and Department of Energy for partial
support during the completion of this work. 

%%%%%%%%%%%%%%%%%%%%%%%%%%%%%%%%%%%%%%%%%%%%%%%%%%%%%%%%%%%%%%%%%%%%%%%%%%%%
%%%%%%%%%%%%%%%%%%%%%%%%%%%%%%%%%%%%%%%%%%%%%%%%%%%%%%%%%%%%%%%%%%%%%%%%%%%%

%%%%%%%%%%%%%%%%%%%%%%%%%%%%%%%%%%%%%%%%%%%%%%%%%%%%%%%%%%%%%%%%%%%%%%%%%%%%
%%%%%%%%%%%%%%%%%%%%%%%%%%%%%%%%%%%%%%%%%%%%%%%%%%%%%%%%%%%%%%%%%%%%%%%%%%%%

\end{document}